\documentstyle[aps,floats]{revtex}
\begin{document}
\renewcommand{\topfraction}{1.0}
\renewcommand{\theequation}{\arabic{section}.\arabic{equation}}
\newcommand{\beq}{\begin{equation}}
\newcommand{\eeq}{\end{equation}}
\newcommand{\beqn}{\begin{eqnarray}}
\newcommand{\eeqn}{\end{eqnarray}}
\newcommand{\slp}{\raise.15ex\hbox{$/$}\kern-.57em\hbox{$\partial$}}
\newcommand{\slA}{\raise.15ex\hbox{$/$}\kern-.57em\hbox{$A$}}
\newcommand{\lnA}{\raise.15ex\hbox{$/$}\kern-.57em\hbox{$A$}}
\newcommand{\slB}{\raise.15ex\hbox{$/$}\kern-.57em\hbox{$B$}}
\newcommand{\bP}{\bar{\Psi}}
\newcommand{\bC}{\bar{\chi}}
 \newcommand{\hs}{\hspace*{0.6cm}}
\twocolumn[\hsize\textwidth\columnwidth\hsize\csname
@twocolumnfalse\endcsname
\title{Equivalence between a bosonic theory and a massive
 non-local Thirring model at Finite Temperature}
 \author{
M.V.Man\'{\i}as and M.L.Trobo}
\address{Departamento de F\'{\i}sica, UNLP, C.C. 67, (1900) La Plata
Argentina \\Consejo Nacional de Investigaciones
Cient\'\i ficas y T\'ecnicas, Argentina.}
\maketitle
\begin{abstract}
\hs Using the path-integral bosonization procedure at Finite Temperature
we study the equivalence between a massive Thirring model with non-local
interaction between currents (NLMT) and a non-local extension of the
sine-Gordon
theory (NLSG). To this end we make a perturbative expansion in the mass
parameter of the NLMT model and in the cosine term of the  NLSG theory in
order
to obtain explicit expressions for the corresponding partition functions.
We conclude that for certain relationship between NLMT and NLSG
potentials both the fermionic and bosonic expansions are equal term by term.
This result constitutes a generalization of Coleman's equivalence at $T=0$,
when considering a Thirring model with bilocal potentials in the interaction
term at Finite Temperature.
The study of this model is relevant in connection with the physics of
strongly correlated systems in one spatial dimension. Indeed, in the
language of
many-body non-relativistic systems, the relativistic mass term can be shown
to represent the introduction of backward-scattering effects.\\
\noindent {\it PACS number(s):} 11.10.Lm - 11.10.Wx - 05.30.Fk 
\end{abstract}
\vskip2pc]
\section{Introduction}
In the context of two dimensional Quantum Field Theories (2D QFT)  the
bosonization procedure
\cite{Stone}, originally developed in the operator language \cite{Low}
\cite{Col},\cite{Mand}, has been also fruitfully implemented in the
path- integral framework \cite{Gam},\cite{Fur}, \cite{Naon}.
 It has been shown to be very useful to
study a great varity of problems in (1+1) dimensions. In particular
the bosonization technique, based in a decoupling change of
path-integral variables, has become a powerful tool for treating
strongly correlated electron systems in 1D. The use of this
technique enables to study some problems, which appear intractable
when formulated in terms of fermions, in an easier way when
formulated in terms of bosonic fields. The Tomonaga-Luttinger liquid theory
(related with a
quantum wire of interacting 1D electrons) \cite{T}, \cite{L},
quantum Hall edge states \cite{Hall} and quantum impurity problems
such as the Kondo effect \cite{Kondo} are some examples in which
the application of the bosonization technique is more
convenient.\\The actual fabrication of the so called quantum wires \cite{Voit}
have renewed the
 interest on low-dimensional field theories, in particular, in the study of 1D
 fermionic gas. This kind of systems may present a deviation from the usual
Fermi
 liquid behavior for which the Fermi surface disappear and the spectrum
contains only
 collective modes. This situation is known as a Luttinger-liquid behavior
\cite{Haldane}.\\
  In a recent series of works
\cite{NLT}, the bosonization procedure in its path integral version
was extended to non-local QFT; in particular, it was applied to a
Thirring-like model
with massless fermions and a non-local (and non-covariant)
interaction between fermionic currents (NLT). For one particular choice of
bilocal potentials this model
displays the same forward scattering processes that are present in
the zero temperature limit of the Tomonaga-Luttinger model \cite{T},
\cite{L}, which
describes a one dimensional gas of highly correlated spinless fermions
interacting through
their density fluctuations.\\
On the other hand, following the functional treatment at finite
temperature \cite{TST}, it has been studied the thermodynamics of the NLT
model \cite{MNTT}, which constitutes a possible starting point to
examine the thermodynamical properties of a Luttinger liquid
through an alternative field-theoretical formulation.\\ In this
paper we extend this path-integral approach to bosonization in
presence of temperature to the case of a NLT model with massive
fermions. From the point of view of the many-body systems, this
relativistic mass term denotes the introduction of
backward-scattering effects \cite{L-E}.\\
The equivalence between the massive Thirring  and the sine-Gordon
models was first derived, using the operational scheme, by Coleman \cite{Col},
who showed that the perturbative series in the mass parameters of the massive
Thirring and the sine-Gordon theories are equal term by term provided
that certain relations hold.
Coleman's proof of the equivalence between both models was also
extended to the finite temperature case \cite{Del}. \\
Recently, following the path-integral procedures applied in Ref.
\cite{Naon} for the local case, it has been shown that the equivalence
still holds for
a certain non-local generalization of these two models at zero
temperature \cite{Kang}.\\
In the present work we show that the non-local massive
Thirring model is equivalent to a certain non-local extension of
the sine-Gordon theory when both models are considered in their
finite temperature version.\\   This paper
is organized as follows: in Section 2 we present the non-local
massive Thirring model at finite temperature. We make a
perturbative expansion in the mass parameter and arrive to a
completely bosonized expression for the generating functional.
We also consider
 a non-local version of the sine-Gordon theory
and trough an expansion in the cosine term we obtain the
corresponding thermodynamical partition function. Comparing term by
term the perturbative series for both models we show that they are equal if a
certain relationship between the NLMT and the NLSG functional of the
potentials is satisfied.Finally, in the last Section, we
summarize the most interesting aspects of our investigations.
\section{Non-local Partition Functions at Finite Temperature}
\setcounter{equation}{0}
\hs In this section we study  two non-local 2D models, the  massive Thirring
model with a non-local interaction between fermionic currents and a
non-local extension
of the Sine-Gordon theory, both at
finite temperature. We use the imaginary time formalism developed
by Bernard \cite{Ber} and Matsubara \cite{Mat} and the
path-integral approach.
\subsection{ Non-local massive Thirring model}
\setcounter{equation}{0}
 We start with the Euclidean action given by
\beqn
{\cal S}& = & \int_{\beta} d^2x \,\bP~(i \slp - m )~ \Psi   \nonumber \\ &
- & \frac{g^2}{2} \int_{\beta} d^2x\,d^2y ~
[J_{\mu}(x)V_{(\mu)}(x-y) J_{\mu}(y)]
\label{1}
\eeqn
where $\int_{\beta} d^2y~$ means $\int_0^{\beta}dy^0 \int dy^1$ and
$\beta=\frac{1}{k_{B}T}$  with $k_{B}$ the Boltzman's constant and
$T$ the temperature.\\
The fermionic currents $J_{\mu}(x) = \bP(x) \gamma_{\mu}\Psi(x)$
 are calculated using $\gamma_{\mu}$ matrices
defined as
\beq
\gamma_{0} = \left( \begin{array}{cc}
0 \;\; 1 \\
1 \;\; 0 \\
\end{array} \right)
\label{1a}
\eeq
\beq
\gamma_1 =\left( \begin{array}{cc}
0 \;\; i\\
-i \;\; 0 \\
\end{array} \right)
\label{1b}
\eeq
and Dirac fermions with $\Psi_{1}$ and $\Psi_{2}$ components
\beq
\Psi =\left( \begin{array}{cc}
\Psi_{1}\\
\Psi_{2} \\
\end{array} \right)
\label{1c}
\eeq
The arbitrary functions $ V_{(\mu)}(x,y) = V_{(\mu)}(\mid x - y \mid)$
represents
the bilocal potentials which describe electron-electron
forward-scattering interactions. Note that no sum over repeated
indices will be implied when a subindex $(\mu)$ is involved.\\
The partition function of the model is given by
\beq
Z = N N_{F}(\beta) \int_{_{\rm antiper}} D\bP \,D\Psi ~ e^{ - {\cal S}}
\label{2}
\eeq
with N an infinite temperature-independent normalization constant and
$N_{F}(\beta)$ a $\beta$-dependent infinite factor to be  determined.\\
 In the case of the fermionic fields the functional integral must be extended
 over the paths with
antiperiodicity conditions in the Euclidean time variable $x^0$
\beqn
\Psi (x^0+\beta, x^1) & = & - \; \Psi(x^0, x^1) \nonumber\\
\bP~ (x^0+\beta, x^1) & = & - \; \bP~(x^0, x^1)
\label{3}
\eeqn
As it is shown in \cite{MNTT}  one can extend
the path-integral approach to non-local bosonization developed in
\cite{NLT} at finite temperature. Indeed, as in the usual, local and
covariant, massless
Thirring model introducing auxiliary vector fields one can remove
the fermionic quartic interaction and express the partition
function in terms of a fermionic determinant.
Let us start splitting the action in the form
\beq
{\cal S} = {\cal S}_{0} + {\cal S}_{{\rm int}}
\label{3a}
\eeq
where $S_0$ contains massive free fermions
\beq
{\cal S}_{0} = \int_{\beta} d^2x\, \bP ~( i \slp -m )~\Psi
\label{3b}
\eeq
and $S_{{\rm int}}$ includes the interaction terms in the form
\beq
{\cal S}_{{\rm int}} = - \frac{g^2}{2} \int_{\beta} d^2x\, J_{\mu}(x)
K_{\mu}(x)
\label{3c}
\eeq
 where we have defined a new current $K_{\mu}(x)$ as
\beq
K_{\mu}(x) = \int_{\beta} d^2y~ V_{(\mu)}(x,y)J_{\mu}(y).
\label{4}
\eeq
Now we introduce a vector field $\tilde{A_{\mu}}$ as follows
\beqn
\exp \left\{\frac{g^2}{2} \int_{\beta} d^2x J_{\mu}(x)
K_{\mu}(x) \right \} & = & \nonumber \\ \int_{_{\rm per}}
D\tilde{A}_{\mu}\, \delta (\tilde{A_{\mu}} -
K_{\mu})
 \exp\left\{\frac{g^2}{2} \int_{\beta}
d^2x \,\tilde{J}_{\mu}\tilde{A}_{\mu}\right\} & &
\label{4a}
\eeqn
and represent the delta functional through a
$\tilde{B}_{\mu}$-field in the form
\beq
\delta(\tilde{A}_{\mu} - K_{\mu}) = \int_{_{\rm per}} D \tilde{B}_{\mu} e^{-
\int_{\beta} d^2x\, \tilde{B}_{\mu} (\tilde{A}_{\mu} - K_{\mu})}
\eeq
We have to impose periodicity conditions for the bosonic
$\tilde{A_{\mu}}$  and $\tilde{B}_{\mu}$ fields over the range $[0,\beta]$.
If we define
\beq
\bar{B}_{\mu}(x)  =  \frac{2}{g^2} \int_{\beta} d^2y~
V_{(\mu)}(y,x)\tilde{B}_{\mu}(y)
\label{7}
\eeq
the fermionic piece of the action results
\beq
{\cal S} = {\cal S}_0 - \frac{g^2}{2} \int_{\beta} d^2x \,J_{\mu}(x)
[ \tilde{A}_{\mu}(x) + \bar{B}_{\mu}(x)]
\label{7a}
\eeq
We can invert eq.(\ref{7}) to obtain
\beq
\tilde{B}_{\mu}(x) = \frac{g^2}{2} \int_{\beta} d^2y~
b_{(\mu)}(y,x)\bar{B}_{\mu}(y).
\label{7b}
\eeq
with the symmetric potentials  $b_{(\mu)}(y,x)$ satisfying
\beq
\int d^2y~ b_{(\mu)}(y,x) V_{(\mu)}(z,y) = \delta^2 (x-z).
\label{8}
\eeq
and make the following change of variables
\beq
\frac{g}{2}(\tilde{A}_{\mu} +\bar{B}_{\mu}) = A_{\mu}
\label{9}
\eeq
\beq
\frac{g}{2}(\tilde{A}_{\mu} - \bar{B}_{\mu}) = B_{\mu},
\label{10}
\eeq
to arrive at the partiton function
\beqn
Z_{ {\rm NLMT}}  & = & N N_{F}(\beta) \int_{_{\rm per}}
D\tilde{A}_{\mu}\,
D\tilde{B}_{\mu} det_{\beta}(i\slp - g~\lnA - m) \nonumber \\
& \times & \exp{\{-S[A,B]\}}
\label{11}
\eeqn
where the bosonic action is given by
\beqn
S[A,B]& = & \frac{1}{2} \int_{(\beta)} d^2x ~
d^2y~b_{(\mu)}(x,y) \nonumber \\ & [ & A_{\mu}(x) A_{\mu}(y) - B_{\mu}(x)
B_{\mu}(y)]
\label{12}
\eeqn
Since the Jacobian associated with the change $(\tilde{A}, \tilde{B})
\rightarrow (A,B)$
is field and temperature-independent we have absorbed it in the
normalization constant $N$.\\
The presence of a temperature-dependent fermionic determinant in
the partition function enables us to apply the non-local
bosonization scheme to the finite temperature case.\\
We see that, after the
change of bosonic variables, the non-locality remains only in the
purely bosonic part of the action. Moreover, as the $B_{\mu}$ field
is completely decoupled from both the $A_{\mu}$-field and the
fermionic ones, its contribution can be factorized and absorbed in
the normalization constant $N_{F}(\beta)$. In addition,
 for repulsive interactions $B_{\mu}$
corresponds to a negative-metric state. This is not a peculiar feature of
neither non-local nor finite-temperature theories. Indeed, the appearance of
negative-metric states was already stressed by Klaiber in his seminal
work on the usual Thirring model \cite{Kla}. In order to have a well defined
Hilbert space, Klaiber had to disregard these fields. Following the same
prescription in this new context, we are naturally
led to include the decoupled $B_{\mu}$-integral in $N_{F}(\beta)$. Indeed,
as it
is habitual in finite-temperature studies, $N_{F}(\beta)$ contains all the
$\beta$-dependent infinite contributions to the partition function. In
order to reproduce the result for the $T \neq 0$
local massive Thirring model \cite{Del} the integral over
ghost-fields should be also a $\beta$-dependent infinite factor, so
it is natural to include it in $N_{F}(\beta)$.\\
Note that, as we stressed before, in eq.(\ref{11}) we arrived at a
partition in which all dependence with non-locality is contained
in the bosonic part. For that reason we can deal with the massive
fermionic determinant following the same procedure as in the local
temperature-dependent case \cite{Del}:
 we perform a chiral expansion with $m$ as perturbative
parameter.\\
Then, we make a chiral change of variables in the fermionic
fields.
\beq
\Psi(x) = e^{g [\gamma_5 \phi(x) + i \eta(x)]}\, \chi(x)
\label{13}
\eeq
\beq
\bP(x) = \bar\chi(x) \,e^{g [\gamma_5 \phi(x) - i\eta(x)]}
\label{14}
\eeq
which $\chi(x)$ satisfying the boundary condition $\chi(x_0, x_1)
= - \chi( x_0 + \beta, x_1)$ and $\phi$ and $\eta$ are bosonic fields in
terms of which we
shall express the $A_{\mu}$ field.\\
 With this change of variables the measure transforms as
\beq
D\bP D\Psi = J_{F}(\phi, \eta) D \bar \chi D\chi
\label{15}
\eeq
The jacobian associated to this change in the fermionic
path-integral measure has been first computed for the $T \neq 0$
case, by Reuter and Dittrich \cite{RD} and gives
\beq
J_{F} = e^{\frac{g^2}{2 \pi}\int_{\beta} d^2x \phi {\Box} \phi }
\label{15'}
\eeq
In 1+1 dimensions it is also possible  to split the gauge field
in a longitudinal plus a transversal component in the following
way
\beq
A_{\mu} = \epsilon_{\mu \nu} \partial_{\nu}\phi - \partial_{\mu}
\eta
\label{16}
\eeq
This change of variables introduces another jacobian in
the path-integral measure give by
\beq
DA_{\mu} = det_{\beta}(-{\Box})~D\phi ~D\eta
\label{17}
\eeq
where $ {\Box} = \partial_{\mu}~ \partial_{\mu}$. Note that this
bosonic determinant is temperature-dependent and hence its
contribution is relevant to the partition function in contrast to
the zero-temperature case where it plays no role and can simply be
absorbed in the normalization constant. \\
Putting all this together we obtain the following partition
function
\beq
Z = N N_{F}(\beta) det_{\beta}(-{\Box}) \int D\bar\chi \, D\chi \,
D\phi\,
D\eta \, e^{- {\cal S}_{{\rm eff}}}
\label{19}
\eeq
where
\beq
{\cal S}_{ {\rm eff}} = {\cal S}_{ {\rm fer}} + {\cal S}_{0 {\rm NLB}}
\label{19a}
\eeq
with
\beq
{\cal S}_{ {\rm fer}} = \int_{\beta} d^2x [\, \bar\chi \,( i \slp - m e^{2g
\gamma_{5} \phi})\, \chi]
\label{19b}
\eeq
and
\beqn
{\cal S}_{0 {\rm NLB}} [\phi, \eta] & = & \frac{g^2}{2 \pi} \int_{\beta} d^2x~
 (\partial_{\mu}\phi)^2 + \frac{1}{2}\int_{\beta} d^2x \,d^2y \nonumber \\
 & \times & [\partial_1 \phi (x)\,
 \,b_{(0)}(x,y)\, \partial_1 \phi (y)  \nonumber \\
 &  + &  \partial_0 \phi (x)\, b_{(1)}(x,y)\, \partial_0 \phi (y) \nonumber \\
 & + &   \partial_0 \eta(x) \, b_{(0)}(x,y)\,\partial_{0} \eta (y)
\nonumber\\
 & + &   \partial_{1} \eta(x)\, b_{(1)}(x,y)\,\partial_1 \eta (y) \nonumber \\
 & - & 2 \,\partial_{0} \eta(x)\, b_{(0)}(x,y)\,\partial_{1} \phi(y)
\nonumber\\
 & + & 2 \,\partial_{1} \eta(x) \, b_{(1)}(x,y) \,\partial_{0} \phi (y)]
 \label{20}
 \eeqn
 which describes a system of two bosonic fields coupled by
 distance-dependent coefficients.\\
In order to study  the generating functional for the massive
non-local Thirring model we make the following expansion  in the mass
parameter $m$.
\beqn
\exp \left\{ m \int_{\beta} d^2x \, \bar\chi e^{2g \gamma_{5} \phi(x)} \chi
\right \} & = & \nonumber \\ \sum_{n=0}^{\infty} \frac{(m)^n}{n!}
\prod_{j=1}^{n}
 \int_{\beta} d^2x_{j} \, \bar\chi(x_j)\, e^{2g \gamma_{5}
\phi(x_j)}\, \chi(x_j)
\label{20a}
\eeqn
Now, it is convenient to write $Z_{ {\rm NLMT}}$ in terms of the thermal
averages $(<
>_{0})$ corresponding to a theory of free massive fermions and
non-local bosons in the form
\beqn
Z_{ {\rm NLMT}}  =  N\, N_{F}(\beta)\, det_{\beta}(-{\Box})\, det_{\beta}(i
\slp)\nonumber \\ \times \int D\phi \, D \eta \, e^{- {\cal S}_{0 {\rm NLB}}}
\sum_{n=0}^{\infty} \frac{(m)^n}{n!} \nonumber \\ \times  <\prod_{j=1}^{n}
\int d^2x_j ~ \bar\chi(x_j) e^{2g \gamma_{5} \phi(x_j)}
\chi(x_j)>_0 & &
\label{21}
\eeqn
We can use the identity
\beqn
\bar\chi(x_{j}) e^{-2g \gamma_{5} \phi(x_{j})}\chi(x_{j}) & = & e^{-2g
\phi(x_{j})}
\bar\chi(x_{j}) \frac{1 + \gamma_{5}}{2} \chi(x_{j}) + \nonumber \\
& &  e^{2g \phi(x_{j})}
\bar\chi(x_{j}) \frac{1 - \gamma_{5}}{2} \chi(x_{j})
\label{22}
\eeqn
and since
\beqn
\bar\chi(x_{j}) \frac{1 + \gamma_{5}}{2} \chi(x_{j}) & = &
\bar\chi_{1}
\chi_{1}(x_{j}) \nonumber \\
\bar\chi(x_{j}) \frac{1 - \gamma_{5}}{2} \chi(x_{j}) & = &
\bar\chi_{2} \chi_{2}(x_{j})
\label{23}
\eeqn
where
\beq
\chi =\left( \begin{array}{cc}
\chi_{1}\\
\chi_{2} \\
\end{array} \right)
\label{24}
\eeq
\beq
\bar\chi =\left( \begin{array}{cc}
\bar\chi_{1}\\
\bar\chi_{2} \\
\end{array} \right)
\label{25}
\eeq
applying the Wick's theorem the eq.(\ref{21}) can be written as
\beqn
Z_{ {\rm NLMT}} & = & N \,N_{F}(\beta)\, det_{\beta}(-{\Box})det_{\beta}(i
\slp) \nonumber \\
 & \times & \int
D\phi\,
D\eta \, e^{-  {\cal S}_{0 {\rm NLB}} }
 \sum _{n=0}^{\infty} \frac{(m)^{2n}}{(n!)^2}  \prod_{j=1}^{n}  \nonumber
\\ & \times & \int
d^2x_{j}\, d^2y_{j}\, < e^{2g \sum_{j}[ \phi(x_{j}) -
\phi(y_{j})]}>_{0 {\rm NLB}} \nonumber \\
& \times &  <\prod_{j=1}^{n} \bar\chi_{1}\chi_{1}(x_{j})\,
\bar\chi_{2}\chi_{2}(y_{j})>_{0 {\rm F}}
\label{26}
\eeqn
in which $< >_{0 {\rm F}}$ means v.e.v in a theory with free fermions and
$< >_{0 {\rm NLB}}$
corresponds to the bosonic action given by eq.(\ref{20}).\\
It is simpler to evaluate the generating functional in the
momentum space. To do this, we Fourier transform the non-local
bosonic action. We expand the bosonic fields $\phi(x^{0}, x^{1})$
and $\eta(x^{0}, x^{1})$ which are periodic in the interval $0 \leq x \leq
\beta$
in a Fourier series
\beq
\phi(x^0,x^1)= \frac{1}{\beta} \sum_{n=-\infty}^{\infty}\int \frac{dk}{2
\pi} \,e^{ik x^1}e^{i\omega_{n}x^0}
\,\tilde{\phi_{n}}(k)
\label{27}
\eeq
where
\beq
\tilde{\phi_{n}}(k)=\int dx^1 \int_{0}^{\beta} dx^0 \, e^{-ik
x^1}e^{-i\omega_{n}x^0}\,
\phi(x^0,x^1)
\label{28}
\eeq
with  $\omega_{n}= \frac{2 \pi n}{\beta}$ the Matsubara frecuencies. A similar
expansion corresponds to $\eta(x^0,x^1)$. For the bilocal potentials we have:
\beqn
b_{(\mu)}(x^1,y^1,x^0,y^0) & = & \frac{1}{\beta}\sum_{n=-\infty}^{\infty}
\int\frac{dk}{2\pi}  \nonumber \\
& \times & e^{ik(x^1- y^1)}e^{i\omega_{n'}(x^0 -y^0)} \,\tilde
{b}_{(\mu)n'}(k).
\label{29}
\eeqn
As it is habitual working at finite temperature, we use
discrete and continous delta functions given by
\beq
\int_0^{\beta} dx^0 \, e^{i(\omega_{n} - \omega_{n'})x^0} = \beta
\delta_{n,n'}
\label{29a}
\eeq
\beq
\int \frac{dx^1}{2 \pi}\, e^{i(k - k')x^1} = \delta( k - k').
\label{29b}
\eeq
and then the Fourier transformed bosonic action results
\beqn
 S_{0 {\rm NLB}}& =& \frac{1}{2\beta}\sum_{n=-\infty}^{\infty} \int
 \frac{dk}{2\pi}\,
[ \tilde{\phi}_{n}(k)\, \tilde{\phi}_{-n}(-k)\,
           A_{n}(k) \nonumber \\
           & + & \tilde{\eta}_{n}(k)\, \tilde{\eta}_{-n}(-k)\,
           B_{n}(k) \nonumber \\ & + & \tilde{\phi}_{n}(k)\,
           \tilde{\eta}_{-n}(-k)\, C_{n}(k) ],
\label{30}
\eeqn
where
 \beq
A_{n}(k) = \frac{g^2}{\pi}( \omega_{n}^2 +  k^2) +
                 k^2~  \tilde{b}_{(0),n}(k) +
           \tilde{b}_{(1),n}(k) \omega_{n}^2
\label{31}
\eeq
\beq
B_{n}(k) = \tilde{b}_{(0),n}(k) \omega_{n}^2 +
\tilde{b}_{(1),n}(k) k^2
\label{32}
\eeq
\beq
C_{n}(k) =2 [\tilde{b}_{(1),n}(k) -
\tilde{b}_{(0),n}(k)] \omega_{n}  k.
\label{33}
\eeq
The bosonic v.e.v  appearing in eq.(\ref{26}) can be written in the form
\beqn
& < & e^{2g \sum_{j} [\phi(x_{j}) -
\phi(y_{j})]}>_{0{\rm NLB}}  =  \frac{1}{{\cal Z}_{0{\rm NLB}}}\int D
\tilde{\phi}(p)\, D \tilde{\eta}(p)
 \nonumber \\  & \times &
\exp \left\{ - \frac{1}{2 \beta} \sum_{n=-\infty}^{\infty}\int \frac{dk}{2
\pi}\, [ \tilde{\phi}_{n}(k)\, A_{n}\, \tilde{\phi}_{-n}(-k) \right.
\nonumber \\
& + & \left.
\tilde{\eta}_{n}(k)\, B_{n}\, \tilde{\eta}_{-n}(-k) +
\tilde{\phi}_{n}(k)\, C_{n}\, \tilde{\eta}_{-n}(-k) \right. \nonumber \\
& - & \left.4 g
\sum_{j=1}^{l} D(\omega_{n},k; x,y)\, \tilde{\phi}_{n}(k)] \right\}
\label{34}
\eeqn
with
\beq
{\cal Z}_{0 {\rm NLB}} = \int D \tilde{\phi}(p)\, D \tilde{\eta}(p)\, e^{-
S_{0 {\rm NLB}}}
\label{34a}
\eeq
and
\beq
D_n(\omega_n,k,x_j,y_j) =  e^{i \omega_n x^0_j + i k x^1_j} - e^{i \omega_n
y^0_j + i k y^1_j}
\label{34cc}
\eeq
We have introduced the vector $p$ which stands for $p=(\omega_{n},k)$.
In order to solve this path-integral we diagonalyze the action in
eq.(\ref{30})
through change of variables
\beqn
\tilde{\phi}_n(k) & = & \phi_n(k) + E_n(k) \nonumber \\
\tilde{\eta}_n(k) & = & \eta_n(k) + F_n(k)
\label{34b}
\eeqn
with
\beqn
E_n(k)& = & 8 g \sum_{n=-\infty}^{\infty} \frac{B_n(k) D_n(k)}{\Delta_n(k)}
\nonumber \\
F_n(k)& = & -4 g \sum_{n=-\infty}^{\infty} \frac{C_n(k) D_n(k)}{\Delta_n(k)}
\label{34c}
\eeqn
and the eq.(\ref{34a}) gets
\beq
{\cal Z}_{0{\rm NLB}}= [det A(k)]^{-1/2} \{det [B(k) - C(k)^2/4
A(k)]\}^{-1/2}
\label{34d}
\eeq
we finally obtain for the bosonic factor
\beqn
& < & e^{2g \sum_{j} [\phi(x_{j}) -
\phi(y_{j})]}>_{0{\rm NLB}}  =   [det A(k)]^{-1/2}\nonumber \\ & \times &
\{det [B(k) - C(k)^2/4
A(k)]\}^{-1/2} \nonumber \\ & \times & \exp \left\{ \frac{- 8 g^2}{\beta}
\sum_{n=-\infty}^{\infty}\int \frac{dk}{2 \pi}
\frac{B_{n}(k)}{\Delta_{n}(k)} \sum_{j,j'}D(\omega_{n},k,
x_{j},y_{j}) \right. \nonumber \\ & & \left.
D(\omega_{-n},-k, x_{j'},y_{j'}) \right\}
\label{34e}
\eeqn
where
\beq
\Delta_{n}(k) = C_{n}^{2}(k) - 4 A_{n}(k) B_{n}(k)
\label{35}
\eeq
The fermionic thermal average is also evaluated giving
\beqn
<\prod_{j=1}^{m} \bar\chi_{1}\chi_{1}(x_{j})\,
\bar\chi_{2}\chi_{2}(y_{j})>_{0{\rm F}}   =
det_{\beta}(i~\slp)\nonumber \\
\times e^{-\frac{1}{\beta}\sum_{n=-\infty}^{\infty}\int \frac{dk}{2
\pi}
\frac{2 \pi}{p^2} D  (\omega_{n},k, x_{j},y_{j})
D(\omega_{-n},-k, x_{j},y_{j})} & &
\label{37}
\eeqn
Putting all this together in eq.(\ref{26}) we obtain for the complete
partition
function
\beqn
Z_{{\rm NLMT}} & = & N \, N_F(\beta) \, det_{\beta}( i \slp) \,
det_{\beta}^{1/2}(-{\Box}) \, det_{\beta}^{-1/2}(\tilde{b}_0 \tilde{b}_1)
\nonumber \\ & \times & det_{\beta}^{-1/2}[\frac{g^2}{\pi}(\frac{
\omega^2_{n}}{\tilde{b}_1} +
\frac{ k^2}{\tilde{b}_0)} + p^2  ] \nonumber \\
 & \times & \sum_{i=0}^{\infty} \frac{m^{2i}}{(i!)^2} \int \prod_{j=1}^{i}
d^2x_{j} \,d^2y_{j}  \nonumber \\ & \times & \exp \left\{
-\frac{1}{\beta}
\sum_{n}\int\frac{dk}{2\pi}
\frac{2 \pi}{\frac{g^2}{\pi}(\frac{\omega^2_n}{\tilde{b_1}} +
\frac{k^2}{\tilde{b_0}}) + p^2} \right. \nonumber \\ & \times & \left.
\sum_{j,j'} D(p, x_{j},y_{j})
D(-p, x_{j},y_{j}) \right\}
\label{38}
\eeqn
where for simplicity we have omitted the $p$-dependence of the
potentials.\\
Thus we have been able to extend an explicit expansion for the
partition function of a massive Thirring model with arbitrary
(symmetric) bilocal potentials coupling the fermionic currents
considered at zero temperature to  the finite
temperature case.\\
\subsection{Non-local extension of the Sine-Gordon model}
Now, we consider  a non-local version of sine-Gordon partition function of the
well known sine-Gordon model. To this end we add in the Lagrangian
density an arbitrary potential function $d_{\mu}(\mid x - y
\mid)$in the form \cite{Kang}
\beqn
{\cal L}_{{\rm NLSG}}&  = & \frac{1}{2} (\partial_{\mu} \Phi)^2 +
\frac{1}{2} \int d^2y \, \partial_{\mu} \Phi(x) \, d_{(\mu)}(x -
y)\,
\partial_{\mu}\Phi(y) \nonumber \\ & - &  \frac{\alpha_0}{\lambda^2} \cos
(\lambda
\phi)
\label{39}
\eeqn
where $\alpha_0$ plays the role of a squared mass and $\lambda$
is a dimmensionless coupling constant.\\
The Euclidean partition function reads
\beq
Z_{{\rm NLSG}} =N_0 \,N'(\beta) \int D\Phi\, e^{- \int d^2x \, {\cal
L}_{{\rm NLSG}}}
\label{40}
\eeq
As in the fermionic case, the renormalization constants $N_0$ and
$N'(\beta)$ absorb the infinite $\beta$-independent and
$\beta$-dependent term respectively. \\
The functional integration runs now over scalar fields periodic in
the time direction
\beq
\Phi(x^0,x^1) = \Phi(x^0 + \beta, x^1)
\label{41}
\eeq
In order to treat this theory we make a perturvative expansion in the
$\alpha_0$ the sine-Gordon partition
function and obtain
\beqn
Z_{{\rm NLSG}} =  N\, N'(\beta)\int D \Phi \, \exp \left\{ \int_{\beta}
d^2x \, \frac{1}{2} (\partial_{\mu}\Phi(x))^2
\right. \nonumber \\   \left. - \frac{1}{2} \int_{\beta} d^2x \, d^2y \,
\partial_{\mu} \Phi(x) d_{(\mu)}(x-y)
\Phi(y)\right\}  \nonumber \\
 \times \sum_{l=0}^{\infty} \frac{1}{(l!)^2}
(\frac{\alpha_0}{\lambda^2})^{2l} \prod_{j=1}^{l} \int d^2x_j \,
d^2y_j
e^{i \lambda \sum_{j}[\Phi(x_j) - \Phi(y_j)]} & &
\label{42}
\eeqn
It is simpler to evaluate the generating functional in the momentum
space. After transforming Fourier the action reads
\beqn
{\cal S}_{{\rm NLSG}}& = & \frac{1}{\beta} \sum_{n} \int \frac{dk}{2
\pi}\, [\frac{(\omega_{n} + k^2)}{2} \tilde{\Phi}_n(k) \tilde{\Phi}_n(-k)
\nonumber \\
& + & \frac{1}{2} (\omega_n^2 \tilde{d}_0(k) + k^2 \tilde{d}_1(k))
\tilde{\Phi}_n(k) \tilde{\Phi}_n(-k)  \nonumber \\
& - & i \lambda \sum_j D_n(\omega_n, k; x_j,y_j)
\tilde{\Phi}_{-n}(k)]
\label{42a}
\eeqn
with
$D_n(\omega_n, k; x_j,y_j)$ given by eq.(\ref{34cc}) which for
simplicity we will denote by $D_n$.\\
Again we solve the path integral in eq.(\ref{42}) translating the
quantum fields $\tilde{\phi}_n(k)$ in the form
\beq
\tilde{\phi}_n(k) = \phi_n(k) + \eta_n(k)
\label{42b}
\eeq
in order to diagonalize the action given by eq.(\ref{42a}). We
arrive at
\beqn
Z_{{\rm NLSG}}& = &N_0 \, N'(\beta) \sum_{l=0}^{\infty} \frac{1}{(l!)^2}
(\frac{\alpha_0}{\lambda^2})^{2l} \nonumber \\ & \times &
det^{-1/2}[\omega^2_n ( 1 + \tilde{d}_0)
+ k^2 ( 1 + \tilde{d}_1)]\nonumber \\
 & \times & \prod_{j=1}^{l} \int  d^2x_j \, d^2y_j  \nonumber \\
& & e^{- \frac{\lambda^2}{2 \beta} \sum_{n}\int \frac{dk}{2 \pi}
\frac{1}{\omega^2_n ( 1 + \tilde{d}_0) + k^2 ( 1 +
\tilde{d}_1)}\sum_{j,j'}D_n D_{-n}}
\label{43}
\eeqn
which constitutes the thermodynamical partition function of the
actual non-local Sine-Gordon model written in terms of arbitrary
potentials $\tilde{d}_{\mu}$.
\subsection{Equivalence between the two non-local models at finite
temperature}
In this section we compare the expressions for the two expansions, one for
the massive non-local Thirring model, eq.(\ref{38}), and the other
for the sine-Gordon like model given by eq.(\ref{43}), both at
finite temperature.
Before doing this let us note that up to now we have considered a
very general situation in which the potentials $b_{(\mu)}$ depend
on both distances and temperature. However from a physical point of
view it is reasonable to assume that the interactions are
temperature-independent. In that case $det^{-1/2}(\tilde{b}_0 \tilde{b}_1)$
which
appears in $Z_{{\rm NLMT}}$ turns to be a $\beta$-independent factor that can
be absorbed in the normalization constant N.\\
On the other hand the finite $\beta$-dependent contributions of
$det_{\beta}^{1/2}(-{\Box})$ and $det_{\beta}(i \slp)$
 cancel each other \cite{Ber}, \cite{Dol} and the infinite
 $\beta$-dependent terms of both determinants fix the value of
$N_{F}(\beta)$ in
 eq.(\ref{38}).\\
 Having this comments in mind we can conclude that the two
 theories we have considered, the  NLTM and the NLSG, are equivalent
provided the
 following relations hold:
\beq
m = \frac{\alpha_0}{\lambda^2}
\label{44}
\eeq
and
\beq
\frac{\lambda^2}{4 \pi ( p^2 + \tilde{d}_0 \omega^2_n +
\tilde{d}_1 k^2)} = \frac{1}{
\frac{g^2}{\pi}(\frac{\omega^2_n}{\tilde{b}_1} + \frac{
k^2}{\tilde{b}_0}
) + p^2}
\label{45}
\eeq
This extension of the Coleman's \cite{Col} equivalence constitutes the main
result of this paper. It is a
generalization to the non-local case of the recently proved
equivalence between sine-Gordon and massive Thirring models at
finite temperature \cite{Del}. To check this result we only have
to take $\tilde{b}_0 = \tilde{b}_1 = 1$ in eq.(\ref{38}) and
$\tilde{d}_0 = \tilde{d}_1 = 0$ in eq.(\ref{43}). These lead us to
\beqn
Z_{{\rm TH}}  =  Z_{{\rm FD}}\, det_{\beta}^{-1/2}( 1+ g^2/ \pi)
\sum_{i=0}^{\infty} \frac{m^{2i}}{(i!)^2} \times \nonumber \\
\int_{\beta} \prod_{j=1}^{i} d^2x_j \, d^2y_j \,
e^{\frac{-1}{\beta} \sum_{n} \int \frac{dk}{2 \pi} \frac{2
\pi}{(1+ g^2/ \pi)} \frac{1}{p^2} \sum_{j,j'} D_n D_{n'}} & &
\label{46}
\eeqn
and
\beqn
Z_{{\rm SG}}  =  Z_{{\rm BE}} \, \sum_{l=0}^{\infty} \frac{1}{(l!)^2}
(\frac{\alpha_0}{\lambda^2})^{2l} \prod_{j=1}^{l} \times \nonumber
\\ \int_{\beta} d^2x_j \, d^2y_j \, e^{\frac{-1}{\beta} \sum_{n}
\int \frac{dk}{2 \pi} \frac{\lambda^2}{2} \frac{1}{p^2}\sum_{j,j'} D_n
D_{n'}}& &
\label{47}
\eeqn
These expressions are the momentum space version of the result
presented in \cite{Del}. Indeed, both expansions are identical
if the following identification is made
\beq
m = \frac{\alpha_0}{\lambda^2}
\label{48}
\eeq
\beq
\frac{4 \pi}{\lambda^2} = 1 + \frac{g^2}{\pi}
\label{49}
\eeq
which coincide with the relations imposed in \cite{Del} when
studying the equivalence between the partition functions corresponding to the
usual massive Thirring model
and the ordinary sine-Gordon model in presence of
temperature and is the well known Coleman's result \cite{Col}.\\Moreover,
in the limit $T \rightarrow 0$ we can also
recover a recent result obtained in \cite{Kang}. In this reference the
authors probed the
equivalence of the two non-local models: the massive Thirring
model and a sine-Gordon one at $T=0$ . To this end they
imposed the same conditions as in the present work, eqs.(\ref{48})
and eq.(\ref{49}), but with the momentum continuous variables $p_0$
instead of the discrete one $\omega_n$
 It is important to stress that eq.(\ref{45}) representing the
 equivalence between both non-local theories is in terms of
 arbitrary potentials. By specifying the interactions it is
 possible to show that these two general theories contain certain
 models already discussed in the literature related to 1-D strong
 correlated systems.
\section{Summary}
\setcounter{equation}{0}
Working in the path-integral formalism at finite temperature we
have shown the equivalence between a non-local massive Thirring
model and a non-local extension of the sine-Gordon theory in
Coleman's sense. To be precise, we have considered a massive Thirring
model with bilocal fermionic current-current interaction and a
sine-Gordon model with a non-local kinetic-like term in its
action.
Following the original procedure developed by Coleman \cite{Col}
but in the present case with the time variable compactified into a
circle of radius $\beta = 1/T$ ( that is at a fixed temperature),
we made perturbative expansions in the mass parameter of the NLMT
model and in the cosine term of the SNSG one. We have shown that
both series are term by term identical provided the relations given
by eqs.(\ref{44}),(\ref{45}) hold. This identities constitutes a
generalization of the Coleman's equivalence for the non-local and
finite temperature case.\\
As it was stressed \cite{NLT} there is a  close relation between a
non-local Thirring model and the physics of strongly correlated
systems in one spatial dimension. For a particular choice of the
bilocal potential corresponding to the case $\tilde{b}_{1} \rightarrow \infty$
and $\tilde{b}_0$ associated to the density-density interaction,
the NLT model coincides with the Tomonaga-Luttinger model, which
constitutes one of the starting points to the study of the 1d
many-body problems. In this kind of systems the presence of a
relativistic fermion mass is related to a Luther-Emery model for
which backward scattering processes are present \cite{L-E}. In
that sense the non-local Sine-Gordon model proposed in this paper
could be the bosonic arena to analyze the thermodynamics of
collective modes in 1d systems when both forward and backward
scattering effects are taken into account.\\
{\bf Acknowledgements:} We would like to thank F.A.Schaposnik and
C.M.Na\'on for a critical reading of the manuscript.\\

\end{document}